\documentclass[seceq]{ptptex}
\usepackage{graphicx}
\usepackage{amsmath,amsthm,amssymb}
\notypesetlogo                       
\newcommand{\PSfig}[2]{\includegraphics[width=#1]{#2}}
\newcommand{\Comments}[1]{}
\usepackage[normalem]{ulem}  
\usepackage[dvips]{color} 

\renewcommand\sout{\bgroup \color{red} \ULdepth=-.5ex \ULset}

\renewcommand{\sout}[1]{}

\newcommand{\ssB}{{\scriptscriptstyle B}}
\newcommand{\ssNN}{{\scriptscriptstyle NN}}
\newcommand{\GeV}{\mathrm{GeV}}
\newcommand{\MeV}{\mathrm{MeV}}

\newcommand{\rhoB}{\rho_{\scriptscriptstyle B}}
\newcommand{\srtNN}{\sqrt{s_\ssNN}}
\newcommand{\muB}{\mu_\ssB}

\newcommand{\TCFO}{T_\mathrm{CFO}}

\newcommand{\muCFO}{\mu_\mathrm{CFO}}
\newcommand{\TCP}{T_\mathrm{CP}}
\newcommand{\muCP}{\mu_\mathrm{CP}}
\newcommand{\horn}{{\em horn}}
\newcommand{\step}{{\em step}}
\newcommand{\dale}{{\em dale}}

\title{%
Phase diagram and heavy-ion collisions: Overview\footnote{
Talk presented at the XLI International Symposium on Multiparticle Dynamics (ISMD2011), Sep. 26-30, 2011, Miyajima, Japan.
Report No.: YITP-11-106.
}
}
\author{
Akira \textsc{Ohnishi}
}
\inst{
Yukawa Institute for Theoretical Physics, Kyoto University,\\
Kyoto 606-8502, Japan\\
}

\abst{
The physics of the QCD phase diagram is discussed
in view from heavy-ion collisions, compact astrophysical phenomena,
lattice QCD and chiral effective models.
We find that $(T,\muB)$ region probed in heavy-ion collisions
and the black hole formation processes
covers most of the critical point locations
predicted in recent lattice QCD Monte-Carlo simulations
and chiral effective models of QCD.
We also discuss the consistency of the statistical model results
and dynamical model description.
}
\begin{document}


\maketitle

\section{Introduction}
While quarks and gluons are the fundamental degrees of freedom
in the Quantum Chromodynamics (QCD),
atomic nuclei are made of nucleons.
When we heat-up or compress nuclei, the QCD phase transition takes place and  
quarks and gluons start to move over a large volume.
Many of the current and past heavy-ion collision experiments have been carried
out to discover the deconfined and thermalized matter
made of quarks and gluons; the quark-gluon plasma (QGP).
Recent experiments at the relativistic heavy-ion collider (RHIC)
and the large hadron collider (LHC)
have shown that strongly interacting matter with high energy density
and characteristic transport properties is created~\cite{Whitepapers},
and this matter is now commonly understood
as a strongly coupled QGP~\cite{sQGP}.
One of the important directions of current and future heavy-ion physics
is to elucidate the properties of QGP
and dynamics of heavy-ion collisions including its preequilibrium stage.
Another intriguing direction is the study of the QCD phase diagram.

\begin{figure}[tb]
\begin{center}
\PSfig{8cm}{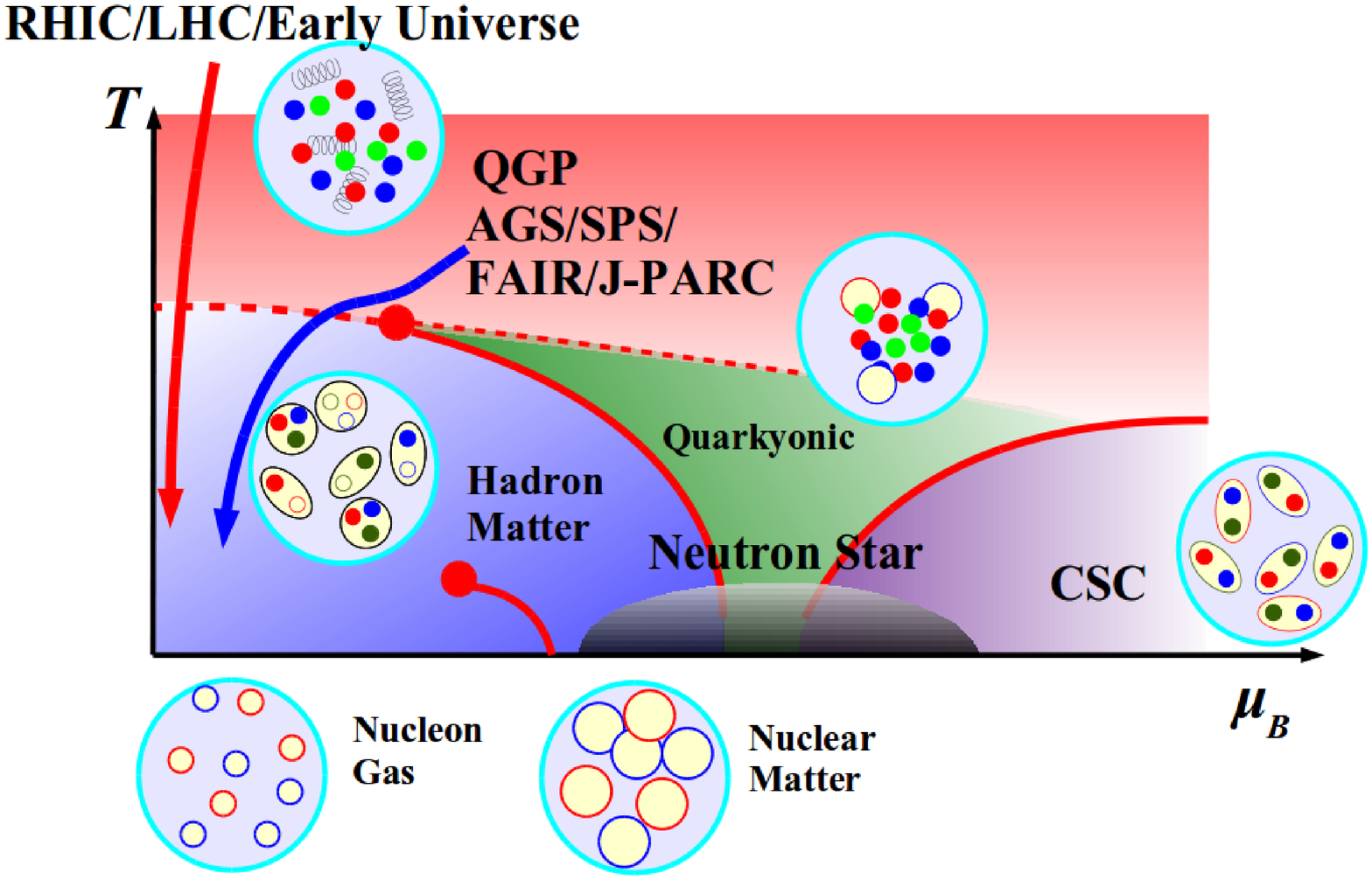}~~~~\PSfig{6cm}{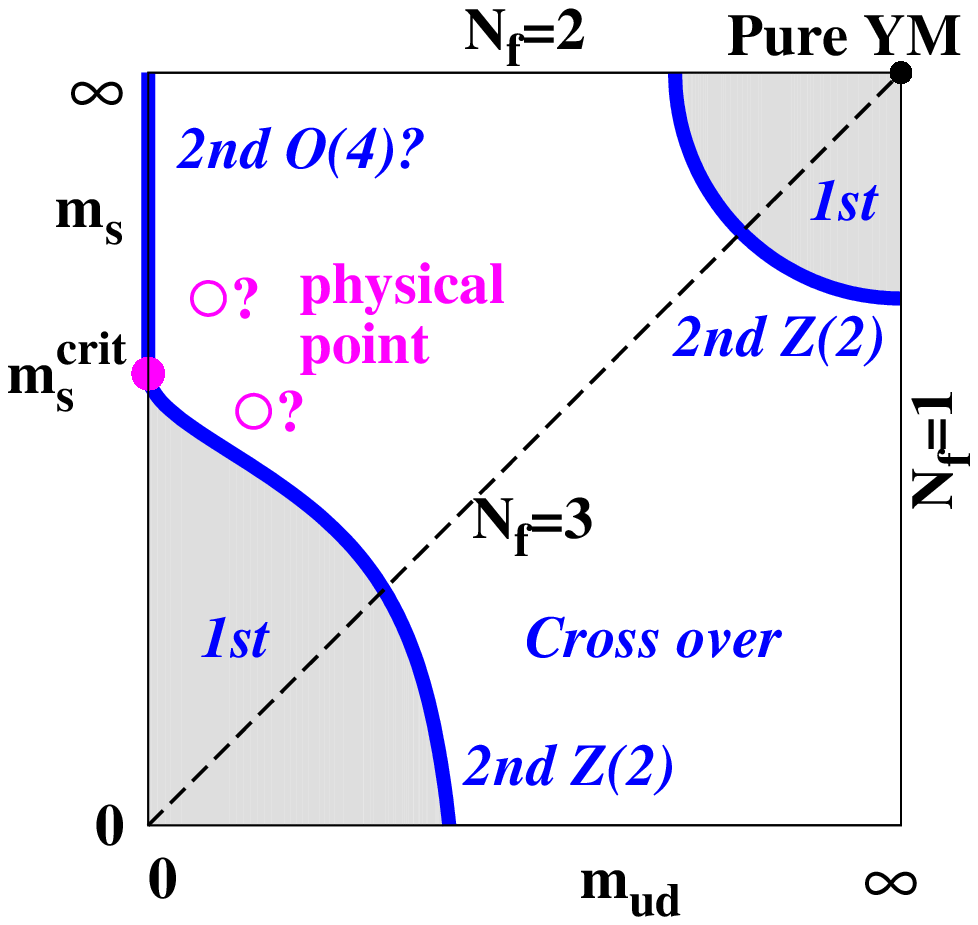}
\end{center}
\caption{
QCD phase diagram (schematic) in $T-\muB$ plane (left)
and in $m_{ud}-m_s$ plane (right).
}
\label{Fig:phase}
\end{figure}

QCD phase diagram has rich structure.
At low temperature ($T$) and low baryo-chemical potential ($\muB$),
the hadron phase is realized where
the chiral symmetry is spontaneously broken and color is confined.
In the very large $\muB$ region,
perturbative QCD predicts that the ground state is the color superconductor
(CSC), where the di-quark pair condensates~\cite{CSC}.
Recent large $N_c$ arguments suggest the existence of another form of matter,
confined high-density matter referred to as the quarkyonic matter~\cite{Quarkyonic}.
These forms of matter can be probed in heavy-ion collisions
and/or compact stars, as shown in Fig.~\ref{Fig:phase}.

\begin{figure}[tb]
\begin{center}
\PSfig{9.5cm}{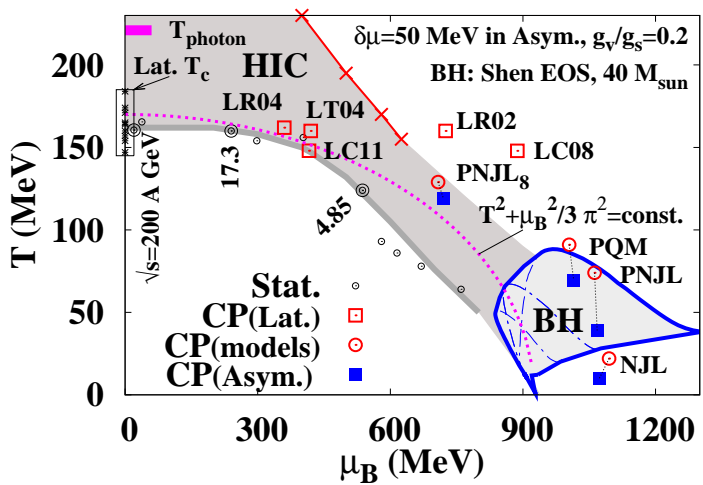}
\end{center}
\caption{
Predictions of the critical temperature and critical point location
in comparison with the chemical freeze-out points, measured photon temperature,
and the swept region during the black hole formation.
}
\label{Fig:CP}
\end{figure}

The QCD phase transition has two facets, deconfinement and chiral transitions.
The deconfinement transition involves a large degrees of freedom change.
In the simplest bag model picture,
pressures in the hadron gas and QGP are given as
$P_\mathrm{H} = 3 P_\mathrm{SB}$
and
$P_\mathrm{QGP} = 37 P_\mathrm{SB} - B$, respectively,
where $P_\mathrm{SB}=\pi^2 T^4/90$
and $B$ is the bag constant.
The degrees of freedom suddenly increases from 3 (pions)
to 37 (quarks and gluons) at $T=T_c=(45 B/17\pi^2)^{1/4} \simeq 0.72 B^{1/4}$,
which gives $T_c \simeq 170~\MeV$ for $B^{1/4}=240~\MeV$.
The spontaneous breaking of the chiral symmetry and its restoration 
are understood as a consequence of the zero point energy
and thermal contribution of quarks.
We consider here a simple model for $N_f=2$.
The fermion contribution to the free energy density
is given as,
\begin{align}
{\Omega_F \over d_f V}
&= - \int^\Lambda {d^3 k \over (2\pi)^3} \frac{E_k}{2}
   - \frac12 \int {d^3 k \over (2\pi)^3} 
	\log (1+e^{-(E_k-\mu)/T})(1+e^{-(E_k+\mu)/T})
\label{Eq:fermion}\\
&= 
	-P_\mathrm{SB}^\mathrm{F} 
	- {M^2 \Lambda^2 \over 16 \pi^2}
	+ {M^2 \mu^2 \over 16 \pi^2}
	+ {M^2 T^2 \over 48}
	+ \mathrm{const.} + {\cal O}(M^4)
\ ,
\end{align}
where $M$ is the fermion mass, $\Lambda$ is the cut-off,  $d_f=4N_cN_f$,
and 
$P_\mathrm{SB}^\mathrm{F}=7/8 \times \pi^2T^4/90 + \mu^2T^2/24 + \mu^4/48\pi^2$.
The first term in Eq.~(\ref{Eq:fermion}) represents the zero point energy,
($E_k/2$ for each momentum $k$),
and the second integral shows thermal contributions.
Provided that the constituent quark mass is proportional
to the chiral condensate $\sigma$
and the free energy density is given as
$\Omega/V=\Omega_F/V + b_\sigma \sigma^2/2$ with a constant coefficient $b_\sigma$,
we find that the second order chiral transition can take place
at $T^2+\muB^2/3\pi^2=T_c^2$ ($\muB=3\mu$),
which is shown by the dotted line in Fig.~\ref{Fig:CP}.
These simple estimates roughly coincide
with the critical temperature $T_c$ obtained in the lattice QCD Monte-Carlo (MC) simulations
and the chemical freeze-out boundary, as discussed later.
The actual QCD phase transition has both of these natures
and is known to be cross over for physical masses
of two light ($u, d$) and one strange ($s$) quarks ($N_f=2+1$)~\cite{CrossOver}
as shown in Fig.~\ref{Fig:phase}.

In this proceedings, we discuss
the QCD phase transition at high $T$ and small baryon densities
in Sec.~\ref{Sec:HighT},
the critical point in Sec.~\ref{Sec:CP},
and the phase diagram structure of dense matter in Sec.~\ref{Sec:Dense}.
We give a short summary in Sec.~\ref{Sec:Summary}.

\section{Phase transition at small densities}\label{Sec:HighT}

In high-energy heavy-ion collisions,
we can probe the transition from QGP to hadron phase
in the high $T$ and low $\muB$ region.
At very high energy, gluons in the small $x$ region governs the dynamics,
and nuclei are regarded as the color-glass condensed (CGC) state.\cite{CGC}
Thermalization takes place after a short time
from the first contact,\cite{thermalization} and QGP is formed.
Collective flows develop
during the hydrodynamical expansion stage,\cite{Hydro,Hirano}
and abundant hadrons are formed at the critical temperature $T\sim T_c$.
Success of hydrodynamics~\cite{Hydro,Hirano}
and statistical models~\cite{Andronic} implies that
local thermal equilibrium is achieved in the early stage at RHIC,
and it enables us to discuss the phase transition in equilibrium.

The cross over nature of the transition at small baryon density
allows us to define $T_c$ in several ways,
such as the chiral susceptibility ($\chi_\sigma$),
strange quark number susceptibility ($\chi_s$),
the Polyakov loop ($P$),
and the $O(N)$ scaling functions ($O(N)$).
The transition temperature in the lattice MC simulation
seems to be converging in the range $T_c=(145\mbox{--}185)~\mathrm{MeV}$.
Around half of the uncertainty comes from the type of fermions.
Simulations with staggered fermions give, for example, 
$T_c(\chi_\sigma)=151(3)(3)~\MeV$
with stout-link improved staggered fermions~\cite{BW2006}
and $T_c(O(N))=157 \pm 6~\MeV$
with asqtad and HISQ/tree actions~\cite{HotQCD2011}.
It is pointed out that the contamination of "heavy" pion
at finite lattice spacing in the previous results~\cite{Cheng2008}
would be the main source of discrepancy.\cite{Kanaya}
Other fermions (domain wall and Wilson fermions)
result in $T_c=(160\mbox{--}184)~\mathrm{MeV}$.\cite{DW-Wilson}
%
I would like to mention that $T_c$ extracted from
the Polyakov loop seems to be systematically higher
than the chiral transition temperature, $T_c(\chi_\sigma)$.
For example
$T_c(P) = 176(3)(4)~\MeV$ is 25 MeV higher than $T_c(\chi_\sigma)$
with stout action.\cite{BW2006}
This ordering supports the idea of deviated chiral-deconfinement transition
boundary at finite $\mu$.\cite{Miura2011}

It is interesting to find that the predicted $T_c$ is close to
the chemical freeze-out temperature in the statistical model fit,
$\TCFO \sim 160~\MeV$~\cite{Andronic}.
The statistical model is based on the assumption that
hadrons are produced in thermal and chemical equilibrium,
and it describes the relative yields of normal hadrons
very well only with two ($T$ and $\muB$) or three parameters.
Approximate agreement of $T_c$ and $\TCFO$ is consistent with
the elliptic flow calculation.\cite{Hirano}
In a hydro+cascade hybrid model,
the rapidity dependence of the elliptic flow ($v_2$) is well explained
when the switching temperature is chosen
to be just below $T_c$.\cite{Hirano,Hydro130}
Elastic scattering dominates in the hadronic cascade stage, 
and chemical composition would be kept approximately.

It should be noted that we cannot determine $T_c$ directly
from $\TCFO$; $\TCFO$ is the temperature in the hadron phase,
and it should be smaller than $T_c$.
In the symposium Gupta proposed a method to evaluate $T_c$
based on data.\cite{Gupta}
By comparing the lattice results with the RHIC-BES fluctuation data
(variance, skewness and kurtosis) systematically,
the critical temperature is evaluated as $T_c=175^{+1}_{-7}~\mathrm{MeV}$.
Another important observation is the direct measurement of the QGP temperature.
Hadrons interact strongly also in the hadronic stage,
then we need direct photon and/or lepton spectra.
PHENIX results of the direct photon spectrum show that
significant excess is observed at $p_T < 3~\mathrm{GeV}/c$
compared to scaled $p+p$ results
and the exponential fit in central collisions gives an apparent inverse slope
parameter of $T_\mathrm{photon}= 221~\mathrm{MeV}$.\cite{PHENIX-photon}
The initial temperature is estimated
to be $T_\mathrm{init}=(300-600)~\mathrm{MeV}$
depending on the thermalization time in hydrodynamical simulations.
These temperatures are clearly higher than $T_c$ estimated in lattice QCD
and $\TCFO$ evaluated in the statistical models,
$T_\mathrm{init} > T_\mathrm{photon} > T_c\ (\mathrm{or}~\TCFO)$.
This ordering provides another evidence of the formation of the QGP
prior to hadronic freeze-out.

\section{Critical Point Search}\label{Sec:CP}

While experiments at RHIC and LHC have been extremely successful
in QGP hunting and revealing the properties of QGP,
the transition at low baryon densities is cross over
and we have not seen evidence of the real (first or second order)
phase transition.
The QCD phase diagram is expected to have a critical point (CP),
which connects the cross over transition at low densities
and the first order phase boundary at high densities
as shown in Fig.~\ref{Fig:phase}.\cite{AsakawaYazaki}
Once the location of CP is identified,
the global structure of the phase diagram is known.
Thus the CP is regarded as the cornerstone of the QCD phase diagram.
Experimental and observational inputs are essential for the CP hunting,
since theoretical predictions of the CP location scatters in a wide region
of the phase diagram as shown in Fig.~\ref{Fig:CP}.\cite{Stephanov}

The lattice MC simulation is the {\em ab initio} framework
to solve non-perturbative QCD,
but it is difficult at present to obtain precise results at high densities
because of the sign problem.
The fermion determinant has the property $D^*(\mu)=D(-\mu)$
and can take a negative value at finite $\mu$.
This determinant is the importance MC sampling weight,
and the cancellation of the weight leads to a large error.
Extensive works have been done or are ongoing to overcome or to avoid
the sign problem.\cite{FodorKatz,LT04,imagmu,LC08,LC11,Nagata,Forcrand-Philipsen,SC-LQCD,deForcrand}.
%
While we have uncertainties from the sign problem,
we find a trend that lattice MC favors high CP temperature $\TCP$
comparable to the zero density critical temperature $T_c(\mu=0)$.
The multi-parameter reweighting method~\cite{FodorKatz}
predicts $(\TCP, \muCP)=(162~\MeV, 360~\MeV)$(LR04),
where $T_c(\mu=0)=164~\MeV$.
Compared to their 2002 results,
$(\TCP, \muCP)=(160~\MeV, 725~\MeV)$(LR02),
smaller quark mass and larger spatial lattice size
result in smaller $\muCP$.
The canonical ensemble method is applicable to a small size lattice,
and a non-monotonic behavior of $\mu/T$ would signal the first order
transition.
The CP location is estimated as
$(\TCP, \muCP) \simeq (0.87\,T_c, 6\,T)$ (LC08)~\cite{LC08}
and 
$(\TCP, \muCP)=(0.927\,T_c, 2.60\,T_c)$ (LC11)~\cite{LC11},
while these results are obtained with heavier $u,d$ quark masses
than physical values.
The Taylor expansion in $\mu/T$ combined with the reweighting
predicts $\muCP \simeq 420~\MeV$ (LT04).\cite{LT04}
On the other hand, the critical mass is found to decrease 
at small non-zero $\mu$ for $N_f=3$,
suggesting there is no chiral critical point.\cite{Forcrand-Philipsen}

Chiral effective models of QCD have been utilized to predict
the phase diagram structure,
including the existence of CP,~\cite{AsakawaYazaki}
and they generally predict the CP location
in lower $T$ and larger $\mu$ region compared with lattice MC results.
For example, NJL~\cite{NJL}, 
Polyakov loop extended NJL (PNJL),\cite{PNJL}
PNJL with 8 quark interaction (PNJL$_8$)~\cite{PNJL8}
and Polyakov loop extended quark meson (PQM)~\cite{PQM} models predict 
$(\TCP,\muCP)=(22,1095), (74,1062), (129, 708)$ and $(91, 1005)$ (in MeV),
respectively.\cite{BHCP}
Here we have taken the vector-scalar coupling ratio of 0.2.

In Fig.~\ref{Fig:CP},
we compare the predicted CP locations
with the chemical freeze-out points at various incident energies
and the $(T,\muB)$ region expected to be probed in heavy-ion collisions. 
We also show the region which would be probed
during the black hole formation processes.\cite{BHCP}
We find that most of the predicted CP locations are
in the accessible region in heavy-ion collisions or black hole formation.
LR02 and LC08 predictions are in the inaccessible region,
but updated or more recent results (LR04 and LC11) are reachable.
Next, a simple estimate of the phase boundary $T^2+\muB^2/3\pi^2=T_c^2$
is found to roughly coincide with
the predicted CP locations and chemical freeze-out line.
We also find that the CP may be close to the chemical freeze-out line
in $\sqrt{s_\ssNN}=(5-130)~\mathrm{GeV}$ range~\cite{Stephanov},
{\em i.e.} between the top AGS energy
($E_\mathrm{inc}=10.6~A~\mathrm{GeV}, \sqrt{s_\ssNN}=4.85~\mathrm{GeV}$)
and the RHIC energy in the first run
($\sqrt{s_\ssNN}=130~\mathrm{GeV}$),
including the top SPS energy
($E_\mathrm{inc}=158~A~\mathrm{GeV}, \sqrt{s_\ssNN}=17.3~\mathrm{GeV}$).
This expectation is consistent with collective flow studies;
AGS energy heavy-ion collisions are well described
in hadron-string cascade models~\cite{Cascade,JAM-RHIC,UrQMD-Hydro},
and the elliptic flow at $\sqrt{s_\ssNN}\geq 130~\mathrm{GeV}$
is successfully described in hydrodynamics~\cite{Hirano,Hydro130}
rather than cascade.\cite{JAM-RHIC}.

The Beam Energy Scan (BES) program at RHIC~\cite{BES} is promising,
since the above incident energy range is fully covered.
As discussed in this symposium~\cite{BES},
the quark number scaling of the elliptic flow
is found to hold at $\srtNN \geq 39~\GeV$,
but not for $\phi$ meson at $\srtNN=11.5~\GeV$.
This scaling appears from the collective dynamics in the partonic stage,
then the above observation may suggest the onset of deconfinement
in the range $11.5~\GeV < \srtNN < 39~\GeV$.
RHIC-STAR collaboration also finds the reduction of the kurtosis
compared with the hadron resonance gas model results
at $\srtNN \lesssim 20~\GeV$.\cite{BES}
The distribution of the the order parameter $X$
follows $\exp[-S_\mathrm{eff}(X)]$,
where $S_\mathrm{eff}$ is the effective action
in which other variables than $X$ are integrated out.
In the vicinity of CP,
$S_\mathrm{eff}(X)$ becomes very flat around the value at CP ($X_c$),
and $X$ will have a distribution $~\exp[-a (X-X_c)^4]$.
This distribution lead to a negative kurtosis $\kappa$,
defined as the correlated part of the 4-th order correlation.\cite{Stephanov}
This large fluctuation may be observed via baryon number fluctuations,
since the chiral condensate couples with the quark number density
at finite $\muB$ and the order parameter becomes a mixture of
the chiral condensate and quark number density.\cite{Fujii}
The sign of the kurtosis or the skewness for the proton number 
along the chemical freeze-out line is sensitive to the model details
and still under debate.\cite{Moments}
We need to understand the behavior of these fluctuation observables
both from theoretical and experimental sides.
Theoretically, we need to go beyond the mean field treatment
and to include fluctuation effects, and it is also important
to understand how strong the baryon number couples with the order parameter.
Experimentally, we need to obtain data with smaller error bars
at smaller interval of incident energies.

Non-monotonic behavior of several quantities is also observed at SPS;
\horn, \step\ and \dale.\cite{NA49-Horn}
The \horn\ structure is the sharply enhanced $K^+/\pi^+$ ratio
observed at $\srtNN \simeq 8~\GeV$, and is proposed as the signal
of QGP formation in statistical model based
on the bag model EOS.\cite{Gazdzicki:1998vd}
The strangeness density is similar or even lower in QGP,
provided that the initial $NN$ collisions make abundant strange hadrons.
Then the $K^+/\pi^+$ ratio will have a sharp peak at $T \sim T_c$,
decrease during the mixed phase, and reach a constant value of QGP.
The \horn\ is also reproduced in a more recent statistical model analysis;
$\TCFO$ is found to saturate (\step) but $\muCFO$ continues to decrease
at $\srtNN \simeq 8~\GeV$,
then $K^+$ having $u$ quark is disfavored at higher energies.\cite{Andronic}
Hadron-string cascade models cannot reproduce \horn,
but explain the \step\ to some extent;
a large part of particles are produced via string fragmentation
at $\srtNN \sim 5~\GeV$,
then suppressed re scatterings during the formation time
leads to the inverse slope saturation.\cite{Cascade,JAM-RHIC}
The description of the \horn\ is improved
in the hadronic cascade+hydrodynamics hybrid model.\cite{UrQMD-Hydro}
The latter three suggest the need of faster thermalization than cascade,
but QGP formation is not necessarily required.
Further studies are necessary to pin down the onset of deconfinement.
Beam energy and system size scan in NA61/Shine may be helpful
for this purpose.\cite{NA61}

\section{Phase structure in dense matter}\label{Sec:Dense}

\begin{figure}[tb]
\begin{center}
\PSfig{7cm}{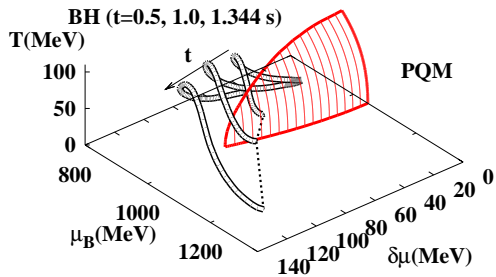}
\end{center}
\caption{
First order phase transition surface in $(T, \muB, \delta\mu)$ space
calculated with PQM is compared with the BH formation profile, 
at $t=0.5, 1.0, 1.344$ sec after the bounce (double lines).
$t=1.344$ sec corresponds to the time just before the black hole formation.
}\label{Fig:BH}
\end{figure}

Phase transition in cold dense matter is important
in compact astrophysical phenomena.
The baryon chemical potential in neutron star core
is calculated to be $\muB \sim 1650~\MeV$ ($\rhoB = 1.12~\mathrm{fm}^{-3}$)
for $M_\mathrm{max}=2.17~M_\odot$ star.\cite{IOTSY}
This chemical potential is much larger than the transition chemical potential
to quark matter in many of the chiral effective models,
$\mu_c = (1000-1110)~\MeV$~\cite{BHCP}.
In supernovae, it is suggested that the transition to quark matter
after the collapse and bounce stage would lead to successful explosion
and may be detected by the second shock peak in anti-neutrino~\cite{SecondPeak}.
One of the recent interesting suggestions is the critical point sweep
during the dynamical black hole formation processes.\cite{BHCP}
The majority of massive stars ($M > (20-25) M_\odot$)
may collapse to black holes without explosions
(faint-supernova),
where very hot ($T \sim 90~\MeV$) and dense ($\rhoB \sim 5 \rho_0$)
matter is formed.\cite{Sumiyoshi}
Since the formed matter is isospin asymmetric,
the isospin chemical potential $\delta\mu=(\mu_n-\mu_p)/2=(\mu_d-\mu_u)/2$
is finite and reduces $\TCP$.
The black hole formation profile $(T, \muB, \delta\mu)$ evaluated
in $\nu$-radiation hydrodynamics~\cite{Sumiyoshi}
is found to go through the critical point in asymmetric matter
obtained in some of the chiral effective models as shown in Fig.~\ref{Fig:BH}.
Thus if the critical point is in the low $T$ and high $\muB$ region,
it may be probed in the black hole formation processes.

There are a variety of proposed phases in cold dense matter,
such as nuclear superfluid, hyperon admixture, meson condensation,
quarkyonic matter, inhomogeneous chiral condensate, color superconductor,
and baryon rich QGP.
The variety of phases is related to the variety of interaction and condensate.
The vector coupling,\cite{Kitazawa2002}
determinant-type six fermi interaction,\cite{KMT}
and eight fermi interaction~\cite{PNJL8,Hiller}
would affect the phase structure at high densities.
When the color superconductor joins the game,
the pairing correlation plays a similar role to temperature
in smoothing the fermi surface,
and we may have another critical point at low $T$.\cite{Kitazawa2002,Kunihiro}

In order to pin down the phase diagram structure in dense matter,
the first principle approaches are of course desired.
One of the promising directions is
the strong coupling lattice QCD,\cite{SC-LQCD}
where we expand the lattice QCD effective potential in $1/g^2$.
Recently, the phase diagram is obtained with finite coupling effects
($1/g^2, 1/g^4$) in the mean field approximation~\cite{SC-LQCD}
and with fluctuation effects in the strong coupling limit.\cite{deForcrand}
Both of these results are obtained in the strong coupling region
($\beta_g = 2N_c/g^2 \lesssim 4$ and $\beta_g=0$)
and based on the unrooted staggered fermion corresponding
to $N_f=4$ in the continuum limit. These are still far from realistic,
but combination of these techniques may be helpful to understand the 
whole structure of the QCD phase diagram.
Another promising direction may be to improve effective models
by using the functional renormalization group approaches~\cite{FRG}.

\section{Summary}\label{Sec:Summary}

In this proceedings, the physics of the QCD phase diagram is reviewed
with emphasis on the relation to heavy-ion collisions.
Lattice MC simulations are powerful for high $T$ and low $\mu$ transition,
and may be useful to discuss the critical point (CP).
The CP location in chiral effective models depends on the model details,
but it roughly follows a simple estimate based on the quark zero point energy
and thermal contribution.
Both hydrodynamics and statistical models are successful at RHIC,
and we can assume local thermalization.
These theoretical estimates of $T_c$ and CP
are compared with statistical model results
and the $(T,\muB)$ region probed in heavy-ion collisions
and black hole formation processes.
If we are not very unlucky, CP would be in the accessible region
in either of two.

As pointed out in the symposium, thermalization may not be achieved
at SPS or lower energies.
Experimentally, it is necessary to cross-check the results
depending on statistical model arguments by using more dynamical observables.
Theoretically, it is challenging to develop theoretical frameworks
which can describe non-equilibrium transport properties 
{\em and} the change of relevant degrees of freedom across the phase boundary.
I could not mention on the chiral magnetic effect
and exotic hadron production in heavy-ion collisions in this proceedings.

\section*{Acknowledgement}

The author would like to thank Prof. A. Nakamura for useful discussions.
This work is supported in part by 
the Global COE Program
"The Next Generation of Physics, Spun from Universality and Emergence",
and the Yukawa International Program for Quark-hadron Sciences (YIPQS).

\end{document}